# Non-uniform Spatial Distribution of Spin Polarization in Topological Insulator ($Bi_2Te_3$) Surface

Ning Wang and Z. Yang

Department of Physics

The Hong Kong University of Science and Technology

Clearwater Bay, Kowloon, Hong Kong, China

**Abstract**

We found, through extensive study of micro reflectance difference spectroscopy images of $Bi_2Te_3$ topological insulator surface, the first clear experimental evidence that there is non-uniform spatial distribution of spin polarization (spin domains) in the surface of the topological insulator. These spin domains involve when they are at different temperatures, in an external magnetic field, and can be driven by electric current. Such surface spin domains in topological insulators have never been predicted by any theory so far.

## 1 Introduction

Topology is a mathematic theory for 'knotting'. Its concept can be applied to the energy-momentum relations (band structures) of electrons in solids, although before 2008 they were considered topologically trivial so there were not much interests in the subject. Around 2008 works started to appear in the literature that showed the non-trivial topology of the band structures in certain materials and structures, such as the edge states of 2-dimensional HgTe/CdTe quantum wells, and later the surface states of certain 3-dimensional solids [1], the surfaces of which are metallic with the conduction band and the valence band touching at the center of the Brillouin zone. Such surface band structure is due to the intrinsic crystal symmetry which leads to particular types of topology in the band structure of the electrons and is in principle immune to defects and other perturbations. Because of that, these solids are called 3-dimensional topological insulators (TI). Among the TI, the $Bi_2Te_3$ has attracted much attention because of its more complex band structure and related spin polarization. The surface band structures have been probed by surface control methods based on a combination of photo-doping and molecular-doping to systematically tune the surface electron density [2]. The snow-flake shape of the conduction band is due to the so-called warping term in the electron Hamiltonian and causes vertical (out of the plane of the surface) spin polarization [3]. Such theoretical prediction has been verified later [4]. The magnitude of the vertical spin polarization, however, is only about 25 % of the in-plane counterpart.

A number of novel properties have been predicted by theory. Effective topological field theory [5] predicts striking topological magneto-electric effect, where an electric field could generate a topological contribution to the magnetization, with a universal constant proportional to the odd multiples of the fine-structure constant $\alpha = e^2/\hbar c$. This would leads to magnetic monopole like image charges induced by a point electric charge near the surface [6]. Surfaces doped with magnetic impurities would lead to rich spin configurations [7], the properties of which are yet to be explored. Chiral states exist on the edge of TI islands [8]. For a slab-like sample with an external magnetic field perpendicular to its surfaces, there are chiral states delocalized on the four side faces carrying both charge and spin currents. The quantized charge Hall effect $\sigma_{xy} = (2n+1)e^2/h$ will coexist with spin Hall effect. The large magnitude of the spin-charge coupling leads to interesting and observable effects in transport properties [9]. It manifests itself in a non-Ohmic contribution to the voltage drop between a ferromagnetic spin-polarized electrode and a nonmagnetic electrode on the TI surface. It can be tuned by applying a gate voltage, which makes it possible to operate the device as a spin transistor, an essential device in spintronics. A device made of TI with two normal metal electrodes and a ferromagnetic island could act as an efficient spin battery with giant output current even at very small microwave power input driving the precession [10].

The first work on the experimental realization of TI was reported in 2009 [11]. Most works so far are concentrated on the angular resolved photoemission spectroscopy [4, 11]. Scanning tunneling microscope images on step reflection of wavefunction showed that backscattering of the topological states by nonmagnetic impurities is completely suppressed, a spectacular manifestation of the time-reversal symmetry which offers a direct proof of the topological nature of the surface states [12]. $Bi_2Te_3$ doped with a few percent of Mn was found to exhibit ferromagnetism with a transition temperature around 10 K [13].

Although the spin of an electron is locked to its momentum [4], at equilibrium an electron with positive momentum and spin will be canceled by another electron with opposite momentum and therefore opposite spin. For any finite surface area the net spin polarization contributed by all the conduction band electrons is therefore expected to be zero. Spin waves could cause local spin polarization fluctuation, but little is further explored [3]. The spin domain structures usually exist only in ferromagnetic materials. However, as there is no doping with Mn or other magnetic impurities, $Bi_2Te_3$ is not expected to be ferromagnetic [13]. And even for Mn doped $Bi_2Te_3$ ferromagnetism only occurs at temperature around 10 K [13], well below the room temperature.

Non-uniform spatial distribution of spin polarization could occur in a T-shaped waveguide at constant electric current flow [14]. But for a TI surface without magnetic impurity, bias electric current, particular topographic structure, and at room temperature, one would not expect any non-uniform spatial distribution of spin polarization.

We have built and tested a magneto-optical microscopic system that can measure both the in-plane and off-plane electron or ion spin polarization near a sample surface [15]. In this paper, we present clear and strong experimental evidence the presence of spin polarization domains in the surface of $Bi_2Te_3$ thin film. Such phenomenon cannot be explained by the present understanding of TI.

## 2  Experiment

The Bi$_2$Te$_3$ thin film (70 nm thick) was grown on (111) semi-insulating GaAs substrates by molecular beam epitaxy using ZnSe as buffer layer. The sample was in the same batch as the ones reported in Ref. 16.

The micro-reflectance-difference spectroscopy (micro-RDS) was used as a surface magneto-optic Kerr Effect (SMOKE) device [15] to study the tiny change of the polarization of the reflected light from the surface of the TI sample. The experimental set-up is the same as we have reported earlier [15], and is shown in Fig. 1. The light from a green diode laser (532 nm) passes through a linear polarizer and is directed to the sample through a beam splitter. A microscope objective focuses the beam to a spot of less than 1 μm × 1 μm on the sample surface. The reflected light passes through a photo-elastic modulator (PEM) with its axis of modulation parallel to the plane of incidence. The PEM is operated with 1/2 wavelength retardation during the experiment. Finally, the beam passes through an analyzer (polarized at 45°) and is focused on a Si-photodiode. The electronic signal from the detector contains three components, namely the DC component which is the normal reflection of the sample surface, the AC component which is at the modulation frequency of the PEM (1f - signal) and is proportional to the imaginary part of the RDS signal, and the AC component at twice the PEM frequency (2f – signal) which is proportional to the real part of the RDS signal, or the Kerr rotation angle in the case of SMOKE. In our experiments the DC and the 2f components were measured by using a standard lock-in amplifier. The sample is mounted on a XY stage (not shown). Usually an 8 μm × 8 μm area was scanned at 0.1μm per step. An aperture was placed between the objective lens and the sample to limit the light beam incidence angle when necessary.

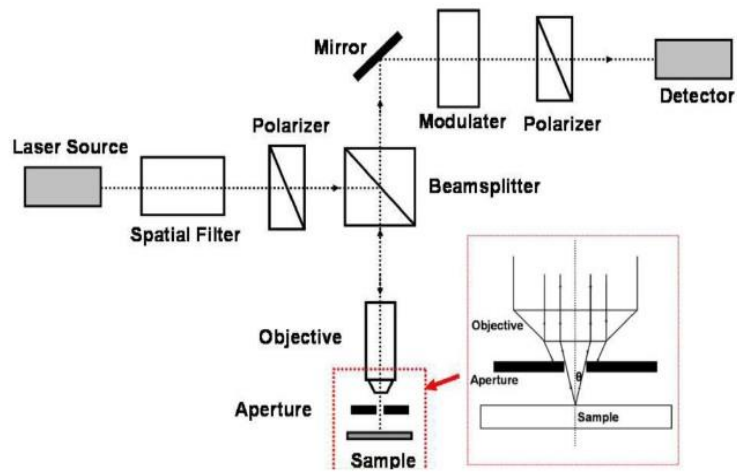

Fig. 1. The schematics of the micro-RDS for the micro-SMOKE experiments.

## 3 Results

There are a number of possible sources contributing to RDS, i. e., the change of light polarization reflected from a sample surface. For a sample without bulk in-plane anisotropy and magnetism, RDS signal could come from four different sources. The first is surface topographic features such as ridges, step edges, or grain boundaries [17]. The second is the anisotropic strain field, especially near the grain boundaries. The third is the conventional anisotropic surface states such as surface dimers. The fourth is non-uniform spatial distribution of surface electron spin density, or surface magnetization domains [14]. The surface topographic features and strain fields do not change when the sample is in an external magnetic field, or at different temperatures. The surface state contribution will not change under different aperture size, is without rotational symmetry, and has little or no dependence on magnetic field. By studying RDS under different conditions we can eliminate certain sources and eventually identify the only source contributing to the observed RDS.

A 2-dimensional RDS image together with the DC (normal reflection) image taken at room temperature is shown in Fig. 2. The unit of the scale bars is milli-volt for the DC image, and $10^{-3}$ for the RDS signal. To be noted are the difference in texture of the two images, and the amplitude variation (minimum-maximum difference) of the images. The DC image (the left image) shows some vague pattern with about 10 % amplitude variation across the 8 μm × 8 μm image, while the RDS signal strength (the right image) varies from 0.2 ($\times 10^{-3}$) to 1.2, a change by 6 times across the same imaged area. Distinct domains can be clearly identified in the RDS image. The sizes vary from a few to a faction of micrometers. As will be shown in the remaining part of this paper, the RDS domains change when the sample is under an external magnetic field, or after several hours of electric current flow, or at different temperatures. Therefore, we can confidently conclude that the RDS images are due to the non-uniform spatial spin polarization.

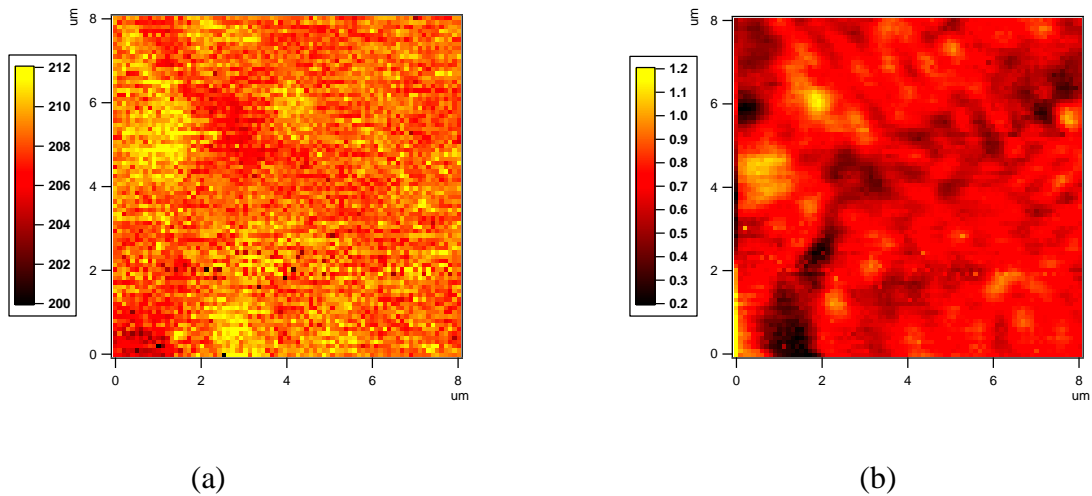

(a)                                         (b)

Fig. 2 Typical normal reflection image (a) and RDS image (b) of the sample at room temperature

## 3.1    In plane/off plane magnetization measurement

In our previous work, we have already demonstrated that even in the pseudo vertical configuration with symmetric illumination one can still detect not only the off-plane (perpendicular to the sample surface) magnetization, but also the in-plane magnetization (parallel to the surface) with good sensitivity [14]. In order to determine whether the magnetization direction of the observed surface spin domains in $Bi_2Te_3$ is in or out of the surface plane, RDS images with different incident aperture sizes were taken. As shown in Ref. 14, the RDS signal strength due to the off-plane spin (or magnetization) is mostly unaffected by the size of the aperture, which blocks off the large angle oblique incident light from the objective lens, while the RDS signal strength due to the in-plane magnetization diminishes when the aperture size is reduced.

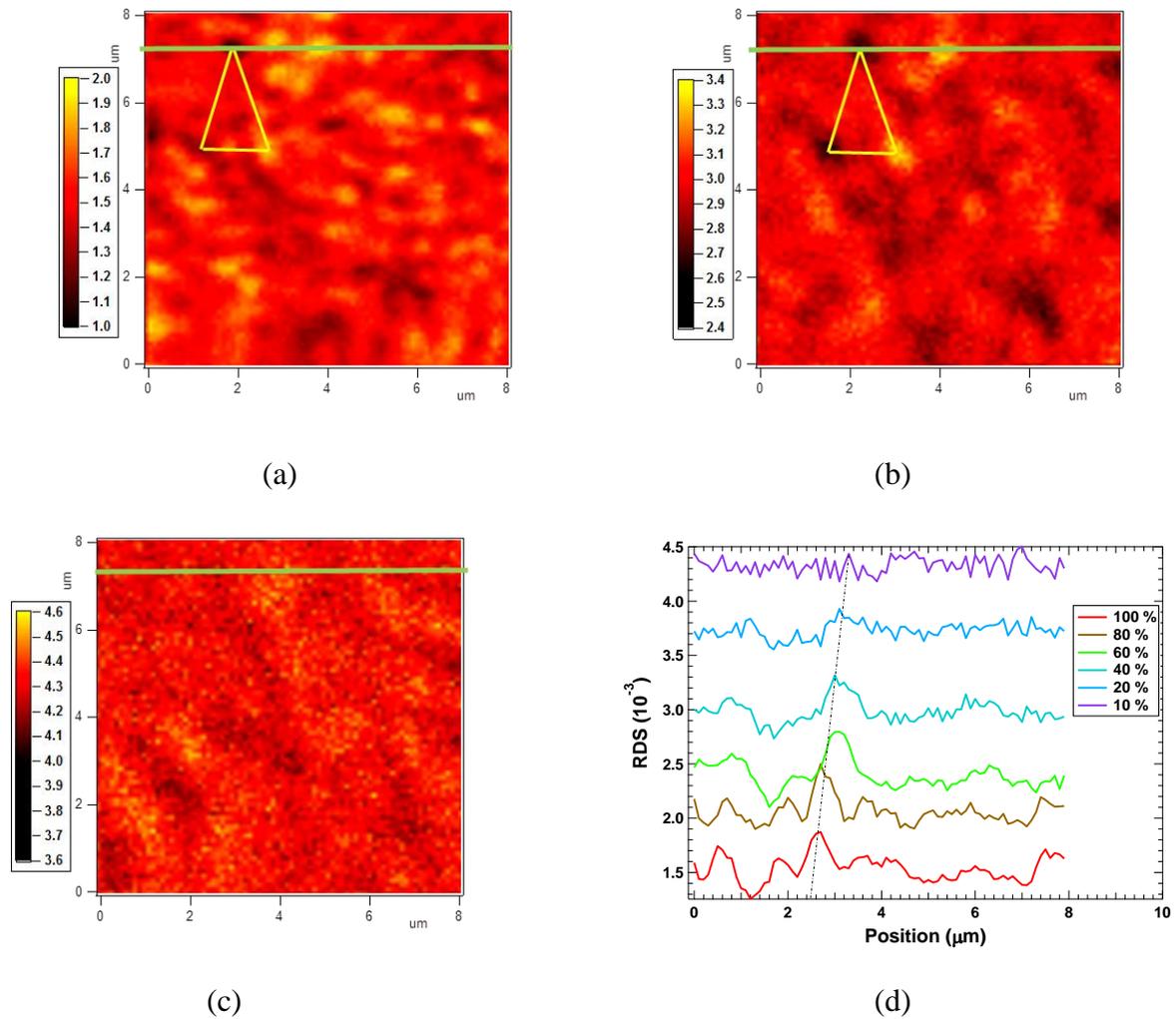

Fig. 3 Micro-RDS images at different aperture sizes (a – c), and a line scan (d). The aperture size

is at (a) 100%, (b) 40%, and (c) 10%, respectively. The green line in (a – c) marks the line scan position.

Typical micro-RDS images at different aperture sizes are shown in Fig. 3. Some 'constellation' pattern as marked in Fig. 3(a) and (b) can be identified to ensure that the same surface area was imaged at different aperture sizes. The contrast patterns become vaguer with decreasing aperture. At 10 % aperture size (Fig. 3(c)) the patterns can no longer be clearly identified. The green line in the images marks the line scan position, and the results are shown in Fig. (d). The marked peak reduces strength with decreasing aperture size, and finally disappears into the noise background at 10 % aperture size. From these results, we can eliminate the possibility of conventional surface electronic states and strain field, and conclude that if the RDS is from the surface magnetization domains, then the direction of the spin in the domains is parallel to the sample surface. However, at this stage we still cannot exclude the possibility of contribution from surface topographic features.

### 3.2   In-plane magnetic field effect

An external magnetic field was applied to the sample to differentiate the last two possible sources of RDS, namely surface topographic features or spin polarization domains. If changes in the domains are observed, it will clearly show that the domains are magnetic in nature.

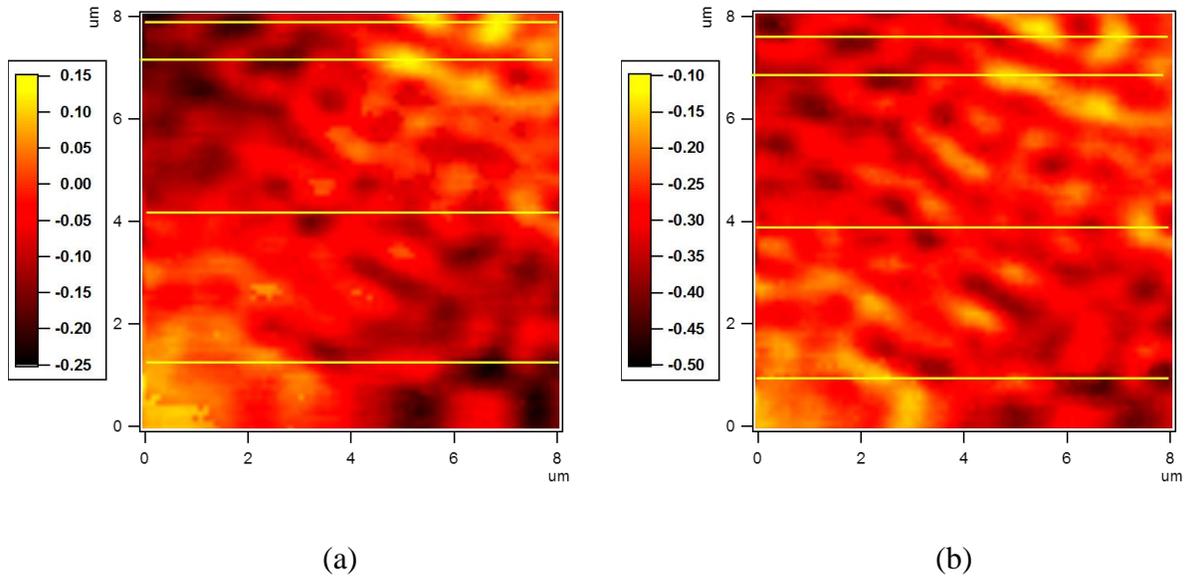

(a)          (b)

Fig. 4 RDS image of $Bi_2Te_3$ in an in-plane magnetic field (a) zero field, (b) 80 mT.

Figure 4 shows the RDS images at zero field and at 80 mT in-plane magnetic field, which was generated by a permanent magnet. Again, some 'constellations' can be identified to ensure that the same area was imaged. The lines in the figures mark the line scan positions, and the outcome is shown in Fig. 5.

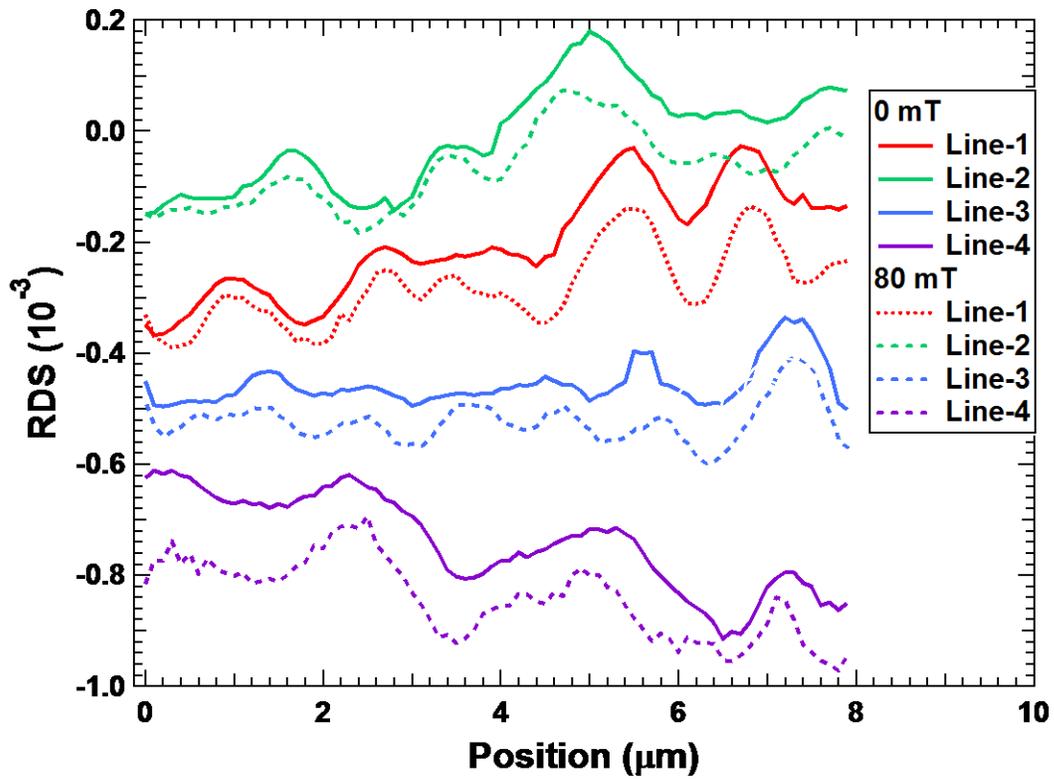

Fig. 5 Line profiles of the RDS images in Fig. 4

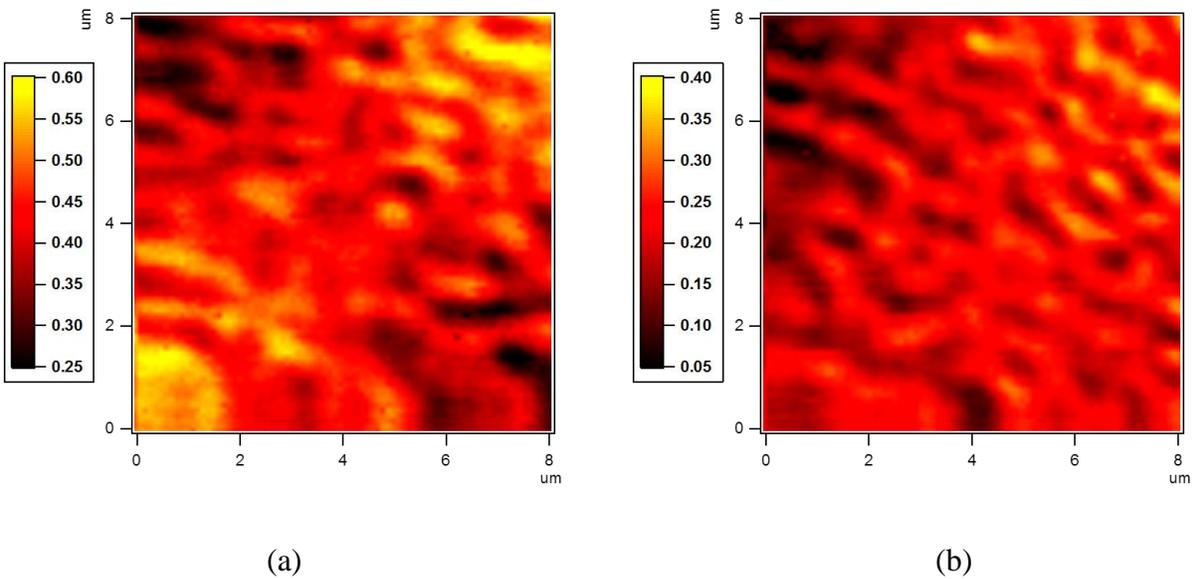

(a)  (b)

Fig. 6 RDS image of $Bi_2Te_3$ in an in-plane magnetic field (a) zero field, (b) 500 mT.

The line profiles shown in Fig. 5 indicate some changes in the domains when the magnetic field was applied. The most significant changes are in line-3 (blue solid and dashed lines). More striking differences can be seen in the images in Fig. 6. While the wavy textures of the images at no field and 500 mT are very similar, one cannot identify any 'constellations' that appear in both images. The change in the domains with external magnetic field is therefore substantial, and there is no doubt that these domains are of magnetic in nature. The possibility that the surface topographic features are the source of RDS can now be safely excluded.

### 3.3 Perpendicular B Field

We have also measured the effect of an external magnetic field which is perpendicular to the sample surface on the spin polarization domains. A circular coil was added in the basic experiment system in order to create a vertical magnetic field. Fig. 7 shows the results of these experiments. A T-shaped 'Constellation' can be identified in the images, as marked by light blue lines. Obvious changes in the domains at different magnetic field strength can be clearly identified. For example, the bar in the T-shaped pattern becomes weaker with increasing magnetic field. The two green lines mark the line scan profiling positions. Only the outcome from the lower line scan is shown in Fig. 6(e). The dip marked by the arrow disappeared when the field was increased to above 40 mT.

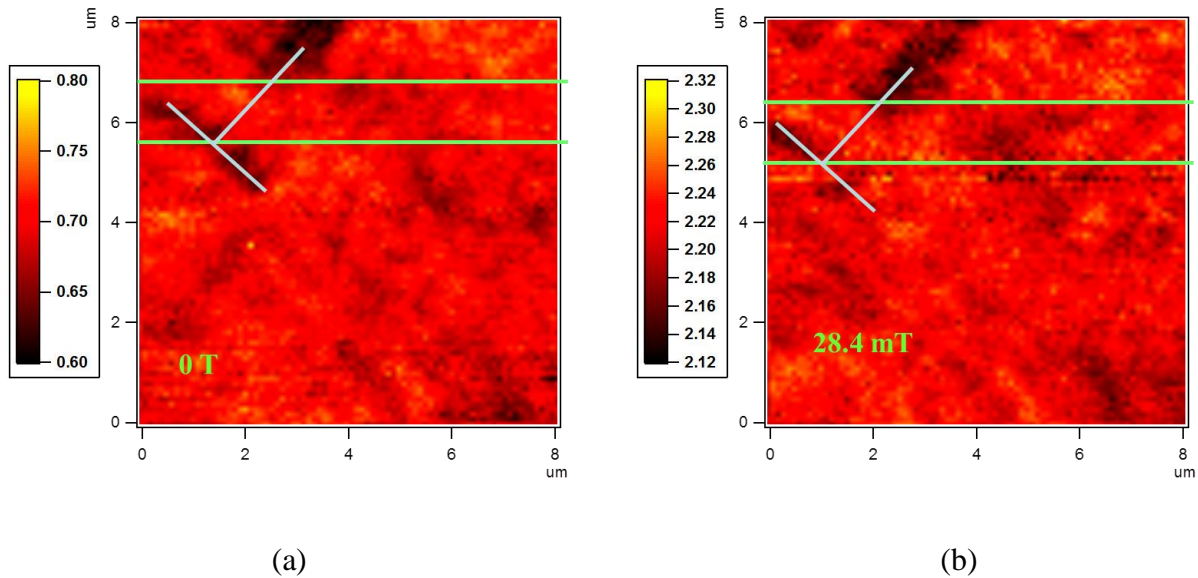

(a)  (b)

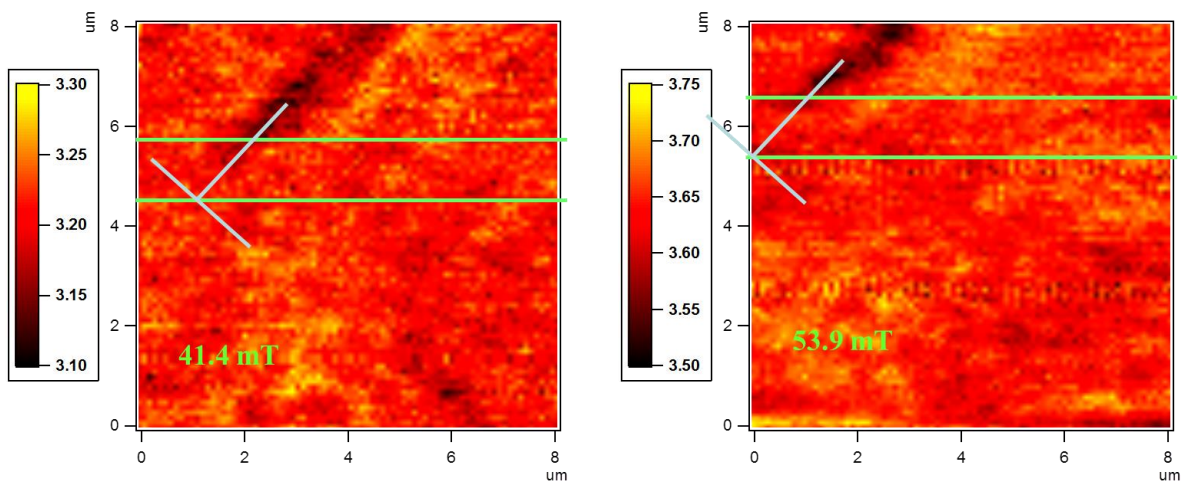

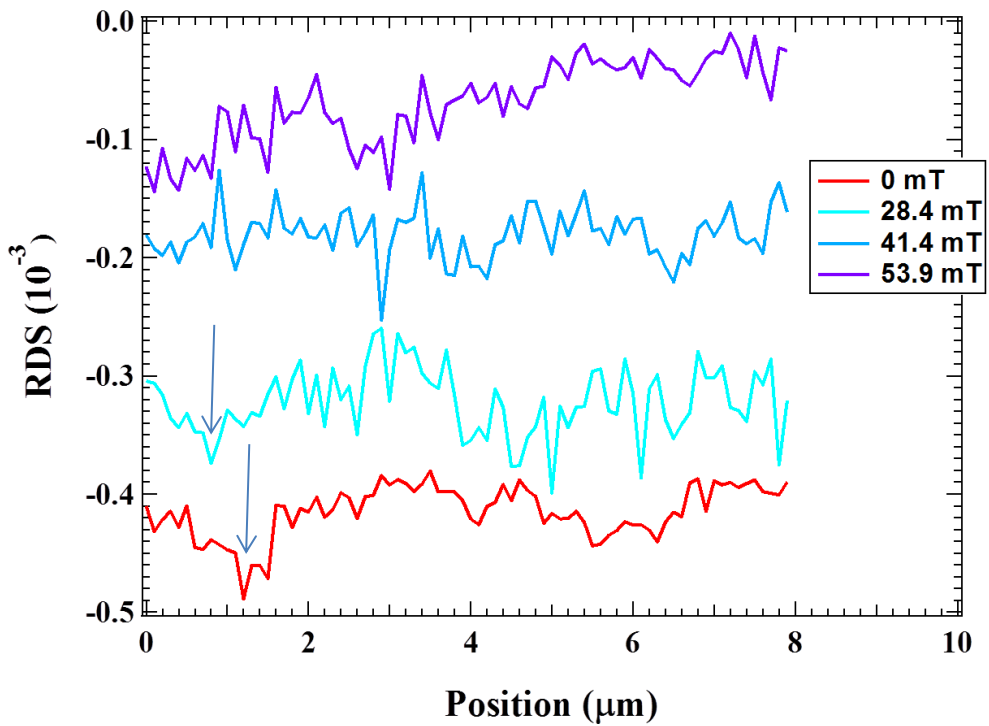

Fig. 7 The RDS images at (a) zero field, (b) 28.4 mT, (c) 41.4 mT, and (d) 53.9 mT, together with the line profile (e).

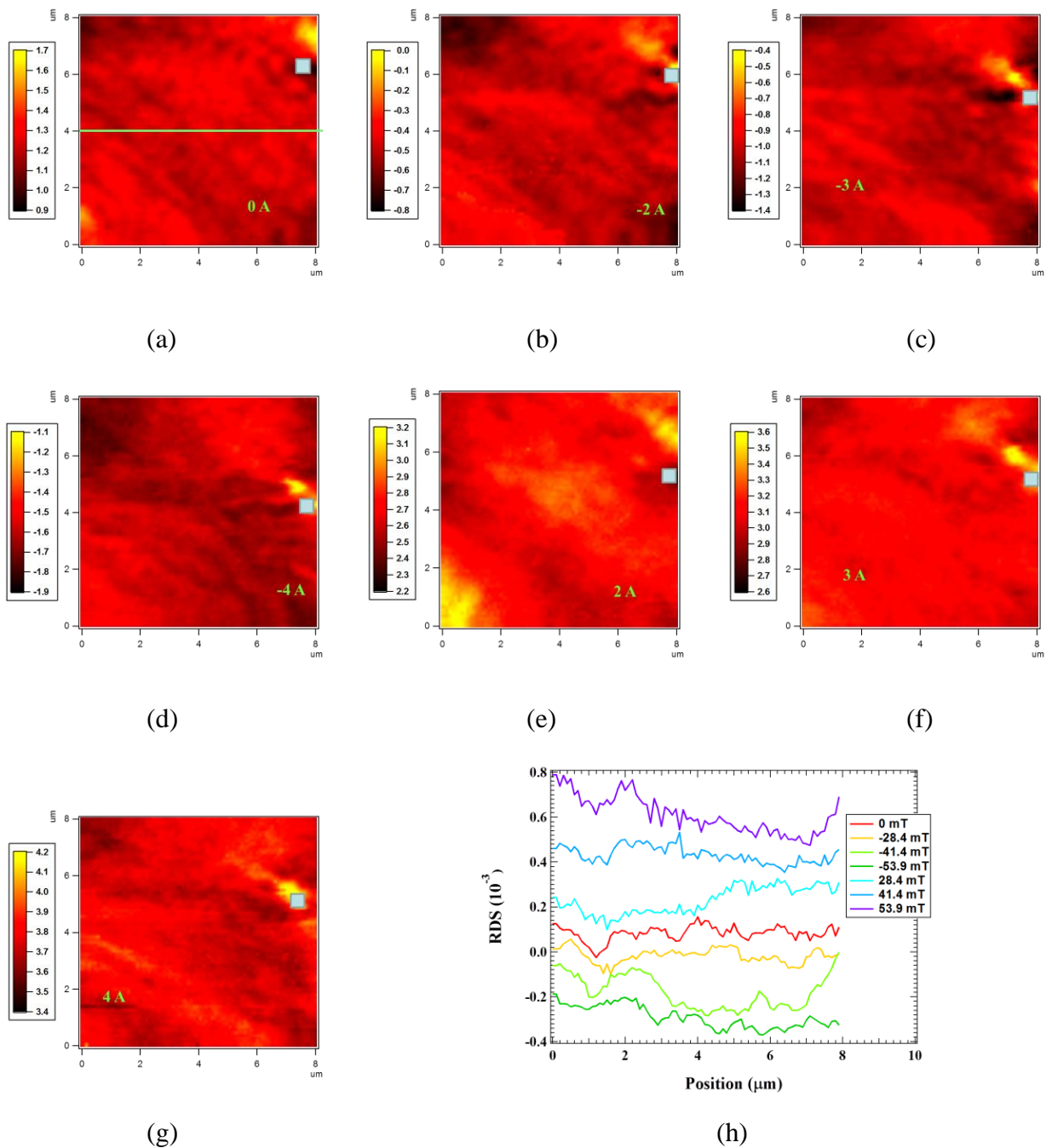

(a) (b) (c)

(d) (e) (f)

(g) (h)

Fig. 8 Another set of RDS images at different magnetic field, including reverse field

Another set of RDS images are shown in Fig. 8. The small square in each image marks the position of a dust particle which happened to be within the area to be imaged. Using this marker, a line scan was taken at fixed distance from the particle, as marked in Fig. 8(a). Again, one can hardly recognize any similar domain patterns among the images at different field. The line profile shown in Fig. 8(h) clearly shows obvious change due to the changing magnetic field.

For comparison, we show in Fig. 9 RDS images of a bare silicon wafer. One can see that there are no domain patterns, and the noise background is no more than 0.1 ($\times 10^{-3}$). Therefore, what we observed and presented above are not due to some artifacts of the micro-RDS system.

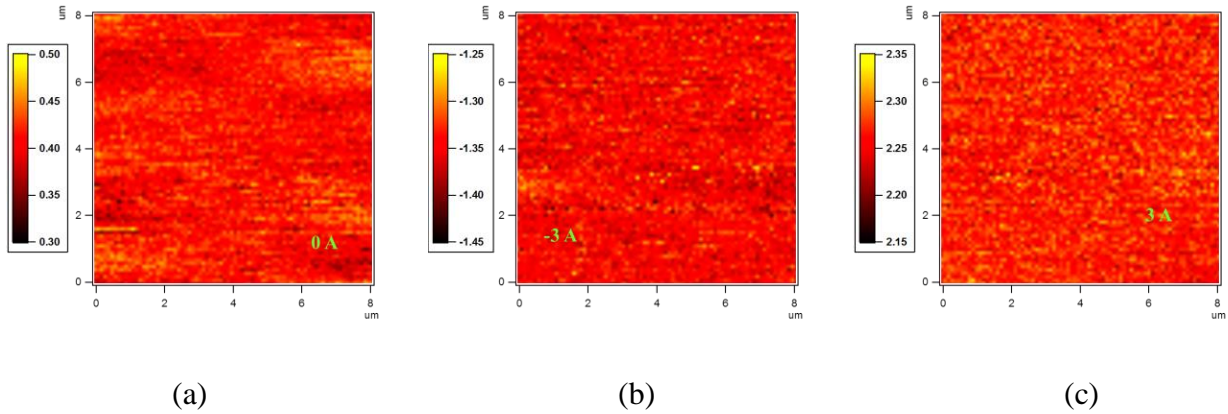

(a)                  (b)                  (c)

Fig. 9  The RDS images of a silicon wafer at (a) zero field, (b) -41.4 mT, and (c) 41.4 mT.

### 3.4 Electric current driven domain evolution

Electric currents are known to drive the magnetic domains in metallic magnetic materials. Through two electrodes near the edges of the sample, DC current at fixed strength was applied for 2 hours and then the images were taken. The current density is roughly $3 \times 10^6$ A/cm$^2$ at 20 mA, as the substrate is insulating. The results are shown in Fig 10. Figure 10(a) is a normal reflection image, where another dust particle (the block patch) can be clearly seen. Then, out of the 15 μm × 15 μm area a smaller square at fixed relative coordinate to the dust particle was taken. The images so taken are shown in Fig. 10(b) for no current, after 10 mA current (Fig. 10(c)), and after 20 mA current (Fig. 10(d)). The line scan position is marked by the yellow line.

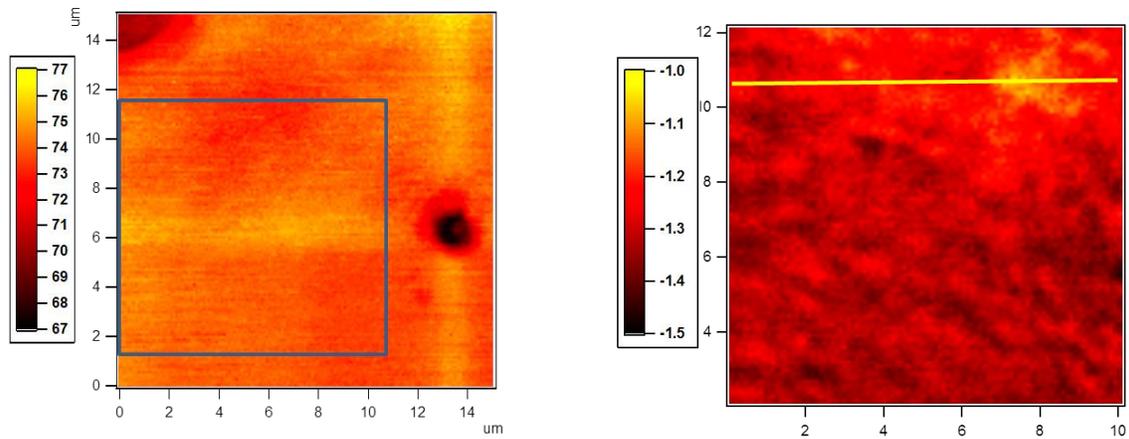

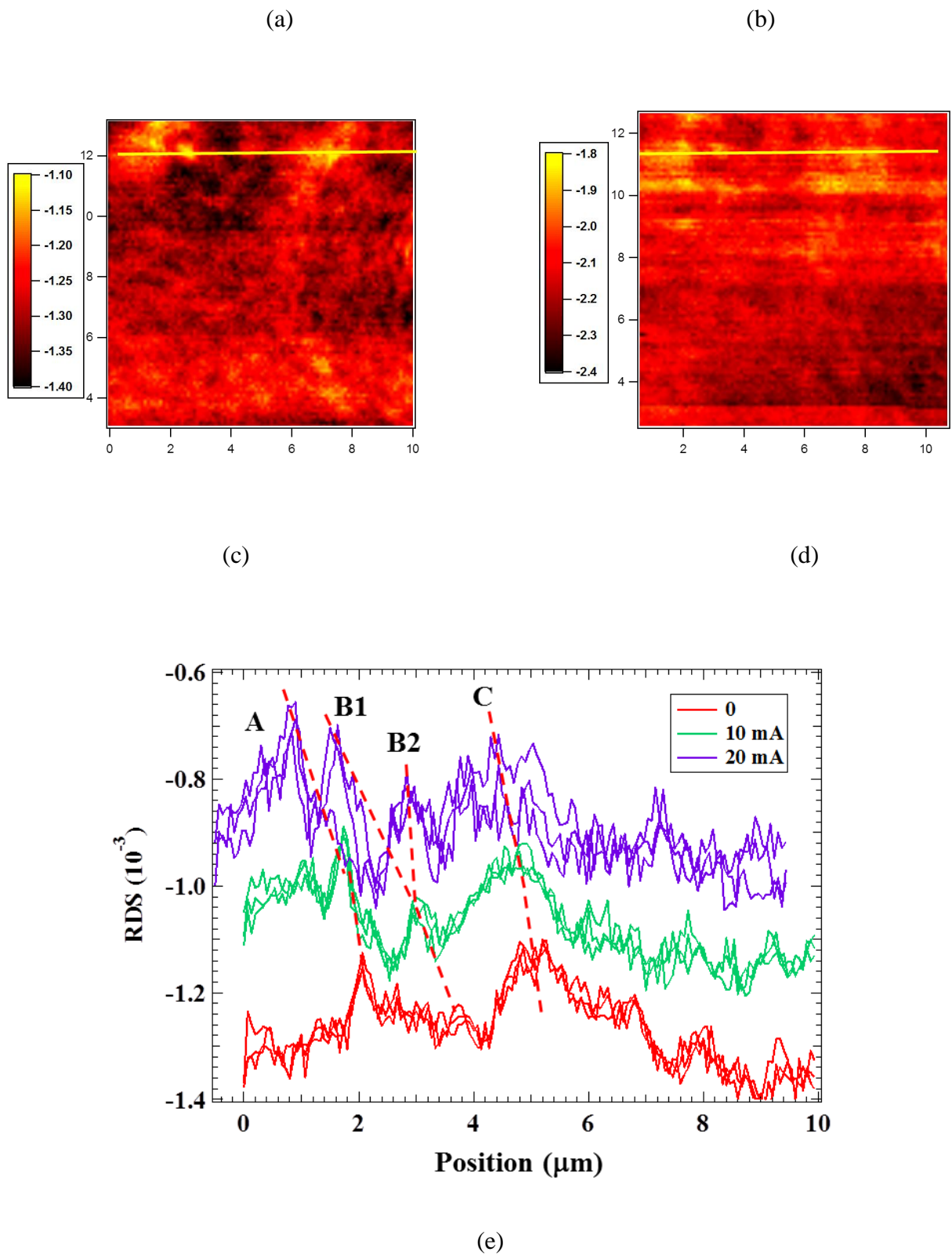

Fig.10 Micro-RDS results of Bi$_2$Te$_3$ with current injection. (a) Normal reflection image showing a dust particle serving as the marker; (b) RDS image before current injection; (c) RDS image

after 2 hours of current at 10 mA; (d) RDS image after 2 hours of current at 20 mA;(e) The line profile scan.

The line profile shows clear shift in position of the spin domains. In general, the domains shifted along the direction of the applied current. For domain-A as marked in Fig. 10(e) the shift distance is about 1 μm. Domain-B, which is very weak before current injection, grew to two strong ones B1 and B2. Domain-C shifted by about 1 μm and while became wider in the process. It is clear that the domains responded to the electric current injection. The current induced spin domain change further strengthen the conclusion that the observed RDS patterns are not due to surface topographic features such as step edges or grain boundaries.

### 3.5 Temperature dependence of spin polarization

Due to thermal drift it is difficult to scan a fixed area over a wide temperature range without visible makers on the sample surface. We therefore used photolithography to deposit an array of photoresist square grids, each about 100 μm in side length on the sample surface. A continuous flow cryostat for optical microscope was used to control the temperature of the sample. The window of the cryostat is made of thin glass plate about 0.2 mm in thickness so as to limit the change of light polarization as the light beam passes through it. When a desired temperature was reached, over 1 hour of holding time was applied to wait for thermal drift to come to stop before each RDS image was taken. The stability of the temperature is better than 1 K. Figure 11 shows a series of images taken at different temperatures in chronicle order. Starting from 300 K (a), the images were taken at 223 K (b), 173 K (c), 123 K (d), and 90 K (e). Then image at 90 K (f) was taken again before the temperature was raised. Subsequently images were taken at 123 K (g), 173 K (h), and finally at 323 K (i) which is above room temperature.

The observed change of spin domains with the change of temperature is dramatic and irreversible. Comparing the first two images taken at 300 K and 223 K, the domain patterns have already changed beyond recognition. Note that it took about 8 hours to take each image, during which the domains could change, as is evidenced by the two images (image-5 in (e) and image-6 in (f)), both taken at 90 K. The contrast (maximum-minimum RDS strength) changed from 2.5 ($\times$ $10^{-3}$) in image-5 to 2.0 in image-6. And the domains patterns are different. The general trend is that at lower temperatures the domains are bigger in size, and the contrast increases also. Apparently the domains were undergoing slow changes when the temperature was changed. After a thermal cycle through low temperature, the domains recover slowly at room temperature to back to the small fragments often observed at room temperature. Such phenomenon adds to the difficulty in the effort to identify a clear pattern of domain evolution dynamics at different temperatures.

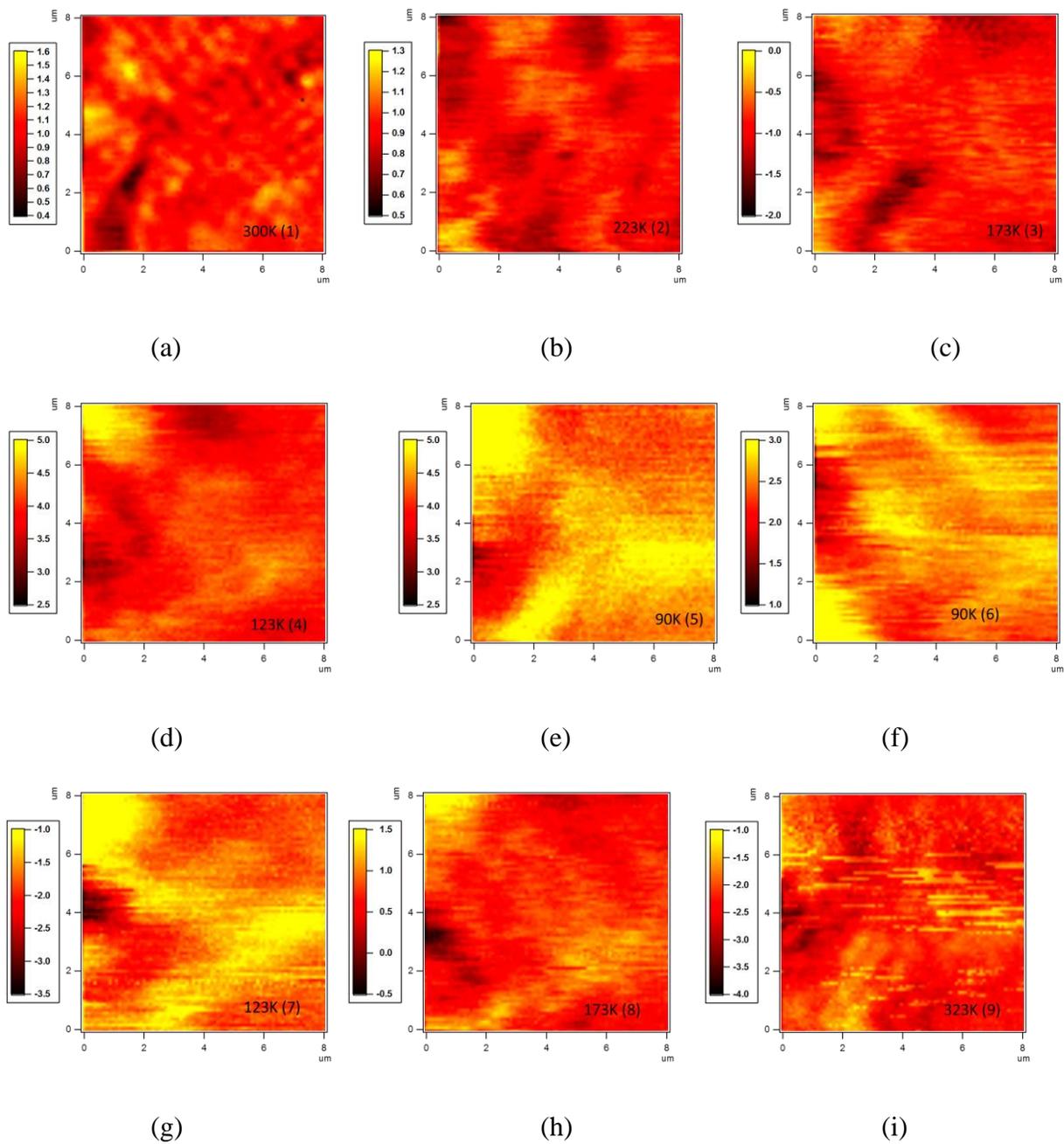

Fig. 11 RDS images in chronicle order at different temperatures. The sequence is labeled by the number in the bracket. For example, 123 K(7) means the image is the 7th in the sequence and was taken when the sample was at 123 K.

## 4   Conclusions

Through extensive study of the micro-RDS images of $Bi_2Te_3$ topological insulator surface, we have obtained clear and strong evidence that the domains in the images are non-uniform spatial distribution of spin polarization, or spin domains. The direction of the spin in the domains is parallel to the surface plane. The domains evolve when the $Bi_2Te_3$ topological insulator sample is in an external magnetic field, at different temperatures, and when a relatively large electric current is passing through. Due to the extremely weak signal, the great difficulty in locating the same scan area especially for the temperature variation experiments, and time limitation (each image took 8 hours), only qualitative or at best semi-quantitative results have been obtained so far. While there is no doubt that spin domains are observed, many questions remain, such as how they behave in magnetic field, temperature, electric current, etc.. The most puzzling, of course, is the underlying physics of such spin polarization domains, which are neither theoretically predicted nor experimentally observed by others so far.

Although the spin of an electron is locked to its momentum [4], at equilibrium an electron with positive momentum and spin will be canceled by another electron with opposite momentum and therefore opposite spin. For any finite surface area the net spin polarization contributed by all the conduction band electrons is therefore expected to be zero. Spin waves could cause local spin polarization fluctuation, but little is further explored [3]. The spin domain structures usually exist only in ferromagnetic materials. However, as there is no doping with Mn or other magnetic impurities, $Bi_2Te_3$ is not expected to be ferromagnetic [13]. And even for Mn doped $Bi_2Te_3$ ferromagnetism only occurs at temperature around 10 K [13], well below the room temperature. Therefore the existing theory in the literature cannot explain what we have observed here. It is the wish of this author that this paper will stimulate more research in the area of spin domains in topological insulators.

Acknowledgement ---- We thank I. K. Sou for providing the sample.